\documentclass[aps,prd,twocolumn]{revtex4}
\usepackage{pdfpages}
\usepackage{graphicx}
\usepackage{amsmath}
\usepackage{booktabs}
\usepackage{multirow}
\usepackage{makecell}
\usepackage{color}
\begin{document}

\title{trimeson state $\boldsymbol{\bar{B}\bar{B}^*\bar{B}^*}$}

\author{Cheng-Rong Deng${\footnote{crdeng@swu.edu.cn}}$ and Chun-Sheng An}
\affiliation{School of Physical Science and Technology,
Southwest University, Chongqing 400715, China}

\begin{abstract}

We systematically explore the trimeson states $\bar{B}\bar{B}^*\bar{B}^*$ with
various isospin-spin configurations in the quark model by solving exactly the six-body
Schr\"{o}dinger equations with the Gaussian expansion method. The configuration
$\left[[\bar{B} \bar{B}^*]^1_0\bar{B}^*\right]^0_{\frac{1}{2}}$ is not only approximately
10.2 MeV lower than the threshold of its constituent particles but also about 0.2
MeV below that of the compact tetraquark state $[\bar{B}\bar{B}^*]^1_0$ and
$\bar{B}^*$. This configuration manifests a loose two-body bound state composed of
$[\bar{B}\bar{B}^*]^1_0$ and $\bar{B}^* $, with a size of around 4.75 fm. In contrast,
the configurations $\left[[\bar{B}\bar{B}^*]^1_1\bar{B}^*\right]^0_{\frac{1}{2}}$,
$\left[\bar{B}[\bar{B}^*\bar{B}^*]^0_1\right]^0_{\frac{1}{2}}$, and
$\left[[\bar{B}\bar{B}^*]^1_1\bar{B}^*\right]^1_{\frac{1}{2}}$ exhibit binding
energies of less than 1 MeV relative to their constituent particles, establishing a
loose three-meson bound state. After coupling three configurations with $\frac{1}{2}0^-$,
the trimeson state with $\frac{1}{2}0^-$ remains a loosely two-body bound state
with a binding energy around 1.5 MeV and a huge size of 2.20 fm, in which the
configuration $\left[[\bar{B}\bar{B}^*]^1_0\bar{B}^*\right]^0_{\frac{1}{2}}$ is
dominant, contributing $80\%$ to the overall probability. Among the four bound
configurations, the $\sigma$-meson exchange plays a decisive role. The meson pair
$[\bar{B}\bar{B}^*]^1_0$, resembling the short-range strong correlated $p$-$n$ pair
in nuclear physics, prevails over other types of meson pairs. The meson pair
$[\bar{B}\bar{B}^*]^1_0$ not only contributes to the binding mechanisms but
also influences the spatial structures of those stable trimeson configurations.

\end{abstract}
\maketitle

\section{introduction}

Recently, the LHCb Collaboration discovered the doubly charmed state
$T^+_{cc}$ with $IJ^P=01^+$ by analyzing the invariant mass spectrum of
$D^0D^0\pi^+$~\cite{LHCb:2021auc,LHCb:2021vvq}. This state is characterized
by a minimal quark configuration of $cc\bar{u}\bar{d}$. The deuteronlike molecular
configuration $DD^*$ is evident from its characteristic size, which is
manifested in the latest reviews~\cite{Chen:2022asf,Huang:2023jec} and
references therein.

In nuclear physics, the Efimov effect predicts that under certain conditions,
three particles can form a bound state even if the pairwise interactions are
weakly attractive or repulsive~\cite{Efimov:1970zz,Naidon:2016dpf}. This phenomenon
is reminiscent of the formation of atomic nuclei from nucleon clusters bound
by the nucleon-nucleon interaction. Similarly, the observation of the $T^+_{cc}$
state naturally raises the question of whether three charmed meson molecules
can exist by adding one charmed meson to the $T^+_{cc}$ state. As a result,
several three-charmed meson systems, such as $DDD^*$, $DD^*D^*$, and $D^*D^*D^*$,
have been studied, with various theoretical models, suggesting the existence
of several shallow bound states~\cite{Wu:2021kbu,Luo:2021ggs,Bayar:2022bnc,Ortega:2024ecy}.
More theoretical predictions on the properties of trimeson states can be found
in the recent literature~\cite{Wu:2022ftm,Liu:2024uxn}.

In general, a larger reduced mass can lower the kinematic energy barrier,
making it easier for multiparticle systems to bind together. Numerous studies
have shown that the doubly bottomed state $T^-_{bb}$ ($bb\bar{u}\bar{d}$)
with $IJ^P=01^+$ can form a deeper bound state compared to the state
$T^+_{cc}$~\cite{Chen:2022asf,Huang:2023jec}. Before the $T^+_{cc}$ state
was observed by the LHCb Collaboration, various three-bottomed meson systems
were investigated to search for potential bound states~\cite{Garcilazo:2018rwu,Ma:2018vhp}.
The compound state $\bar{B}\bar{B}^*\bar{B}^*$-$\bar{B}^*\bar{B}^*\bar{B}^*$ with
$\frac{1}{2}2^-$ can form a deep bound state, with its binding energy approximately
90 MeV below the threshold for three bottom mesons~\cite{Garcilazo:2018rwu}.
In contrast, the state $BBB^*$ with $IJ^P=\frac{1}{2}1^-$ is a loosely bound
state, exhibiting a binding energy of only a few MeVs in the one-pion exchange
potential model~\cite{Ma:2018vhp}.

The trimeson states have typically been studied under the assumption of a three-body
system at the hadron level using Faddeev equations~\cite{MartinezTorres:2011gjk,Garcilazo:2018rwu}.
A widely utilized and simpler approach is the Fixed Center approximation, which is applied
when two of the three particles are more strongly correlated with each
other~\cite{Bayar:2015zba,Dias:2017miz,Ren:2018pcd,Dias:2018iuy}. Additionally, various
methods such as the Gaussian expansion method~\cite{Wu:2019vsy,Wu:2020job},
coupled-channel complex-scaling method~\cite{Shinmura:2019nqw}, Born-Oppenheimer
approximation~\cite{Ma:2018vhp,Ma:2017ery}, and QCD sum rules~\cite{MartinezTorres:2012jr}
have also been employed to investigate the three-meson system at the hadron level.

Short-range correlations (SRCs) play a crucial role in describing strongly interacting
many-body quantum systems. The SRCs between pairs of nucleons are essential for a
comprehensive understanding of both the fundamental aspects of nuclear dynamics and
the short-range characteristics of the nuclear force~\cite{Arrington:2011xs,Hen:2016kwk,Dalal:2022zkg}.
Among nucleon-nucleon SRCs, the proton-neutron ($p$-$n$) correlations, which share
the same quantum numbers as the deuteron, are significantly more prevalent than the
proton-proton ($p$-$p$) and neutron-neutron ($n$-$n$)
correlations~\cite{CLAS:2018xvc,CLAS:2019vsb,JeffersonLabHallA:2020wrr,Patsyuk:2021fju,Li:2022fhh}.
This raises an important question: do meson pairs exhibit correlations similar to those
seen in nucleon-nucleon SRCs within few-meson systems, and what would be their characteristics?
This topic deserves further theoretical investigation.

In this study, we aim to systematically investigate the trimeson system $\bar{B}\bar{B}^*\bar{B}^*$
at the quark level using the quark model. We will solve the six-body Schr\"{o}dinger
equations exactly employing the Gaussian expansion method~\cite{Hiyama:2003cu,Hiyama:2018ivm}.
Our objective is to explore potential trimeson bound states across various isospin-spin
configurations. We will calculate their binding energies, analyze their spatial configurations,
examine the correlations of meson pairs, and investigate the underlying binding mechanisms
from the perspective of the quark model. We hope to provide valuable insights that could
assist in the experimental identification of three bottom meson states in the future.

After the Introduction, the paper is organized as follows. In Sec. II, we give the
necessary details of the quark model. In Sec. III, we briefly introduce the trial wave
functions for mesons, dimeson states $\bar{B}^{(*)}\bar{B}^*$ and trimeson states
$\bar{B}\bar{B}^*\bar{B}^*$. In Sec. IV, we discuss the natures of the dimeson
states $\bar{B}^{(*)}\bar{B}^* $. In Sec. V we carefully analyse the properties
of the trimeson states $\bar{B}\bar{B}^*\bar{B}^*$. In the last section we list
a brief summary and outlook.

\section{quark model}

The {\it ab initio} calculation of hadron spectroscopy and hadron-hadron interactions
directly from quantum chromodynamics (QCD) is quite challenging due to the
complex nonperturbative nature of the theory. Consequently, the QCD-inspired
constituent quark model is a valuable tool for gaining physical insights into
these complicated strong-interacting systems. This model is based on the premise
that hadrons are color-singlet, nonrelativistic bound states of constituent quarks,
characterized by phenomenological effective masses and interactions.

It is expected that the dynamics of the model are governed by QCD. The perturbative
effect is primarily represented by the well-known one-gluon exchange (OGE) interaction.
From the nonrelativistic reduction of the OGE diagram in QCD for point-like quarks,
we obtain the following expression,
\begin{eqnarray}
V_{ij}^{\rm oge}={\frac{\alpha_{s}}{4}}\boldsymbol{\lambda}^c_{i}
\cdot\boldsymbol{\lambda}_{j}^c\left({\frac{1}{r_{ij}}}-
{\frac{2\pi\delta(\mathbf{r}_{ij})\boldsymbol{\sigma}_{i}\cdot
\boldsymbol{\sigma}_{j}}{3m_im_j}}\right),
\end{eqnarray}
where $\boldsymbol{\lambda}^c_{i}$ and $\boldsymbol{\sigma}_{i}$ represent
the color $SU(3)$ Gell-Mann matrices and the spin $SU(2)$ Pauli matrices,
respectively. $r_{ij}$ denotes the distance between quarks $i$ and $j$,
and $m_i$ is the mass of the $i$-th quark. The OGE interaction is not only
responsible for the mass splitting observed in hadron spectra~\cite{Isgur:1978xj,Isgur:1979be},
but it also contributes to the short-range repulsive core by the spin-spin
part of the inter quark interaction in the nucleon-nucleon interactions~\cite{Liberman:1977qs}.

The constituent quark mass arises from the breaking of the $SU(3)_L\otimes SU(3)_R$
chiral symmetry at a certain momentum scale~\cite{Manohar:1983md}. In this context,
the chiral symmetry is spontaneously broken in the light quark sector, while it is
explicitly broken in the heavy quark sector. Once the light constituent quark mass
is generated, they must interact through Goldstone bosons $\pi$, $K$ and $\eta$.
The explicit Goldstone boson exchange potentials $V_{ij}^{\pi}$, $V_{ij}^{K}$, and
$V_{ij}^{\eta}$, are adopted from Ref.~\cite{Vijande:2004he},
\begin{eqnarray}
\begin{aligned}
V_{ij}^{\rm obe}= & V^{\pi}_{ij}\sum_{k=1}^3\boldsymbol{F}_i^k
\boldsymbol{F}_j^k+V^{K}_{ij}\sum_{k=4}^7\boldsymbol{F}_i^k\boldsymbol{F}_j^k\\
+&V^{\eta}_{ij} (\boldsymbol{F}^8_i\boldsymbol{F}^8_j\cos\theta_P
-\sin\theta_P),\\
V^{\chi}_{ij}= &
\frac{g^2_{ch}}{4\pi}\frac{m^3_{\chi}}{12m_im_j}
\frac{\Lambda^{2}_{\chi}}{\Lambda^{2}_{\chi}-m_{\chi}^2}
\mathbf{\sigma}_{i}\cdot
\mathbf{\sigma}_{j} \\
\times &\left( Y(m_\chi r_{ij})-
\frac{\Lambda^{3}_{\chi}}{m_{\chi}^3}Y(\Lambda_{\chi} r_{ij})
\right),~x=\pi, K~\mbox{and}~\eta.
\end{aligned}
\end{eqnarray}
The scalar meson $\sigma$ exchange interaction is also involved in the model~\cite{Vijande:2004he},
\begin{eqnarray}
\begin{aligned}
V^{\sigma}_{ij}=&-\frac{g^2_{ch}}{4\pi}
\frac{\Lambda^{2}_{\sigma}m_{\sigma}}{\Lambda^{2}_{\sigma}-m_{\sigma}^2}
\left(Y(m_\sigma r_{ij})-
\frac{\Lambda_{\sigma}}{m_{\sigma}}Y(\Lambda_{\sigma}r_{ij})
\right). \\
\end{aligned}
\end{eqnarray}
The $\pi$ exchange interaction and $\sigma$ exchange interaction account for long- and medium-range
behaviors of nuclear force~\cite{Obukhovsky:1990tx,Fernandez:1993hx,Valcarce:1995dm}, respectively.
Additionally, the $K$ exchange interaction is employed to investigate the nucleon-hyperon and
hyperon-hyperon interactions~\cite{Fujiwara:1996qj}.

The quark confinement is one of the most distinctive and important features of QCD.
At present it is still impossible for us to derive the quark confinement analytically
from the QCD Lagrangian. In phenomenological potential models, various types of
potentials are employed to describe the effects of quark confinement~\cite{M:2023hms}.

In the constituent quark models, the quark confinement potential depends on the color charges
of quarks. The coupling between color charges increases with increasing separation between quarks.
Based on the two ingredients, the quark confinement potential can therefore be written as
\begin{eqnarray}
\begin{aligned}
V_{ij}^{\rm con}=-a_c\boldsymbol{\lambda}^c_{i}
\cdot\boldsymbol{\lambda}^c_{j}r^n_{ij}.\\
\end{aligned}
\end{eqnarray}
Some models adopted a linear potential~\cite{Eichten:1974af,Silvestre-Brac:1985aip,Garcilazo:2001md}
while others employed a quadratic one~\cite{Isgur:1978xj,Straub:1988mz,Wang:1992wi}.
Comparative studies of multiquark systems indicate that the distinction between models using
quadratic and linear confinement potentials can be effectively mitigated by adjusting the model
parameters, making the difference less pronounced~\cite{Ping:1998si,Huang:2018rpb}. In the dynamical
calculations of multiquark states, the use of a quadratic confinement potential can significantly
simplify the computational workload, particularly when dealing with the multi-body confinement
potential arising from the color-flux tube in lattice QCD~\cite{Deng:2014gqa,Wang:2023jqs}.
Building on the reasons discussed above and our prior work~\cite{Deng:2022cld,Deng:2021gnb,Lin:2023gzm},
we adopt the quadratic confinement potential in this manuscript.

To sum up, the complete model Hamiltonian for conventional mesons, dimeson states $T^+_{cc}$ and
$\bar{B}^{(*)}\bar{B}^*$, and trimeson states $\bar{B}\bar{B}^*\bar{B}^*$ can be presented as
\begin{eqnarray}
\begin{aligned}
H_n=&\sum_{i=1}^n\left(m_i+\frac{\mathbf{p}_i^2}{2m_i}\right)-T_{c}+\sum_{i<j}^n V_{ij}\\
V_{ij}=&V_{ij}^{\rm oge}+V_{ij}^{\rm con}+V_{ij}^{\rm obe}+V_{ij}^{\sigma},
\end{aligned}
\end{eqnarray}
where $\mathbf{p}_i$ represents the momentum of the $i$-th quark and $T_{c}$ denotes the center-of-mass
kinetic energy.

\section{wave functions}

The wave function of a colorless bottomed meson, denoted as $\bar{B}^{(*)}$, with isospin
$I$ and angular momentum $J$ can be expressed as the direct product of its constituent
parts: the color part $\chi_c$, the isospin part $\eta_{i}$, the spin part $\psi_s$,
and the spatial part $\phi(\mathbf{r})$. This can be written mathematically as
\begin{eqnarray}
\Phi^{\bar{B}^{(*)}}_{IJ}=\chi_c\otimes\eta_{i}\otimes\psi_s\otimes\phi(\mathbf{r}),
\end{eqnarray}
where $\mathbf{r}$ is the relative coordinate between the quarks $b$ and $\bar{q}$,
with $\bar{q}$ representing either $\bar{u}$ or $\bar{d}$ in the $\bar{B}^{(*)}$ meson.

The wave function of dimeson ground states $\bar{B}^{(*)}\bar{B}^*$, denoted as
$T_{bb}$, can be established by the colorless mesons $\bar{B}^{(*)}$ (composed of
$b_1\bar{q}_1$) and $\bar{B}^{*}$ (composed of $b_2\bar{q}_2$). The wave function
of the states with defined isospin $I_{12}$ and total angular momentum $J_{12}$
can be expressed as
\begin{eqnarray}
\begin{aligned}
\Psi^{T_{bb}}_{I_{12}J_{12}}=\sum_{\xi}
c_{\xi}\mathcal{A}_{12}\left\{\left[\Phi^{\bar{B}^{(*)}}_{I_1J_1}
\Phi^{\bar{B}^{*}}_{I_2J_2}\right]_{I_{12}}^{J_{12}}\phi(\boldsymbol{\rho})\right\}.
\end{aligned}
\end{eqnarray}
The square brackets indicate the Clebsch-Gordan couplings of the angular momentum and isospin.
The operator $\mathcal{A}_{12}$ serves as an antisymmetrization operator that acts on the identical
quarks $b_1$ and $b_2$, as well as on the identical anti-quarks $\bar{q}_1$ and $\bar{q}_2$.
\begin{eqnarray}
\begin{aligned}
&\mathcal{A}_{12}=P_{b_1b_2}P_{\bar{q}_1\bar{q}_2},\\
\noalign{\smallskip}
&P_{b_1b_2}=1-P_{b_1b_2},~P_{\bar{q}_1\bar{q}_2}=1-P_{\bar{q}_1\bar{q}_2}.
\end{aligned}
\end{eqnarray}
Here, $P$ is the permutation operator acting on the identical particles. The summation index
$\xi$ encompasses all possible isospin-spin intermediate configurations $\{I_1, I_2, J_1, J_2\}$
that can be coupled to yield the total quantum numbers of the state $T_{bb}$. The coefficients
$c_{\xi}$ can be determined through the dynamics of the model. The term $\phi(\boldsymbol{\rho})$
represents the wave function for the relative motion between the two mesons in the center-of-mass
frame. The Jaccobi coordinate $\boldsymbol{\rho}$ can be explicitly expressed as
\begin{eqnarray}
\begin{aligned}
&\boldsymbol{\rho}=\frac{m_b\mathbf{r}_{b_1}+m_q\mathbf{r}_{\bar{q}_1}}{m_b+m_{\bar{q}}}
-\frac{m_b\mathbf{r}_{b_2}+m_q\mathbf{r}_{\bar{q}_2}}{m_b+m_{\bar{q}}}.
\end{aligned}
\end{eqnarray}

It is worth emphasizing that the primary reason of selecting the meson-meson configuration
$[b\bar{q}][b\bar{q}]$ instead of diquark-antidiquark configuration $[bb][\bar{q}\bar{q}]$
as the starting point is that this manuscript focus on the properties of trimeson state
$\bar{B}\bar{B}^*\bar{B}^*$ and correlations between two meson $\bar{B}^{(*)}$ and
$\bar{B}^*$. The study of the meson-meson configuration $[b\bar{q}][b\bar{q}]$ just serves
as a foundation for further exploration of the trimeson state $\bar{B}\bar{B}^*\bar{B}^*$.

\begin{figure} [h]
\resizebox{0.46\textwidth}{!}{\includegraphics{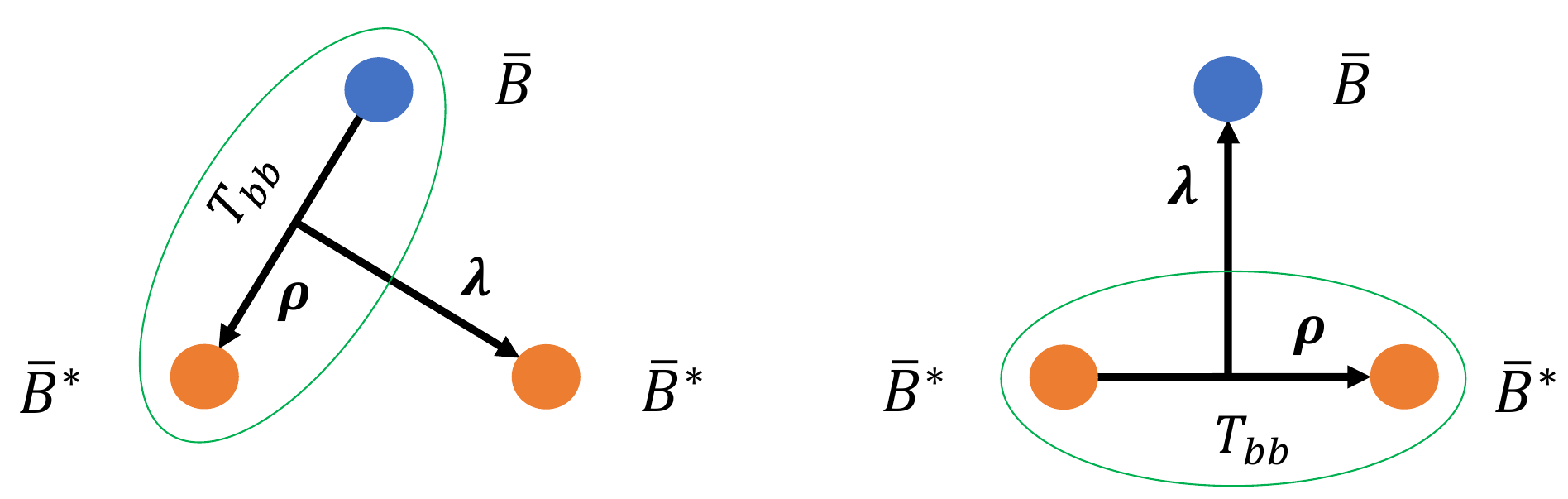}}
\caption{Jaccobi coordinates for the trimeson state $\bar{B}\bar{B}^*\bar{B}^*$. $\bar{B}$
or $\bar{B}^{(*)}$ and $\bar{B}^*$ first forms a dimeson state $T_{bb}$, which then combines
with the spectator $\bar{B}^*$ or $\bar{B}$ to establish a trimeson state.}
\label{jaccobi}
\end{figure}

For the trimeson states $\bar{B} \bar{B}^*\bar{B}^*$, denoted as $H_{bbb}$, we introduce
two sets of Jacobi coordinates, as illustrated in Fig.~\ref{jaccobi}. The left set is used to
describe the correlation between $\bar{B}$ and $\bar{B}^*$, while the right set is employed
to depict the correlation between $\bar{B}^*$ and $\bar{B}^*$ in the trimeson state
$\bar{B}\bar{B}^*\bar{B}^*$.

The procedures for constructing the wave functions using the two sets of Jacobi coordinates
are identical. Taking the left set as an example, the wave functions can be obtained by combining
the well-defined dimeson state $\bar{B} \bar{B}^* $ with the spectator state $\bar{B}^{*}$
(composed of $b_3\bar{q}_3$).

The total wave function of the trimeson states with defined isospin-spin can be expressed as
\begin{eqnarray}
\begin{aligned}
\Psi^{H_{bbb}}_{IJ}=\sum_{\xi}
c_{\xi}\mathcal{A}_{123}\left\{\left[\Psi^{T_{bb}}_{I_{12}J_{12}}\Phi^{\bar{B}^{*}}_{I_3J_3}\right]_I^J
\phi(\boldsymbol{\lambda})\right\}.
\end{aligned}
\end{eqnarray}
$\mathcal{A}_{123}$ is antisymmetrization operator acting on the identical $b$-quarks $b_1b_2b_3$ and
$\bar{u}$- or $\bar{d}$-quarks $\bar{q}_1\bar{q}_2\bar{q}_3$,
\begin{eqnarray}
\begin{aligned}
&\mathcal{A}_{123}=P_{b_1b_2b_3}P_{\bar{q}_1\bar{q}_2\bar{q}_3},\\
\noalign{\smallskip}
&P_{b_1b_2b_3}=1-P_{b_1b_3}-P_{b_2b_3},\\
\noalign{\smallskip}
&P_{\bar{q}_1\bar{q}_2\bar{q}_3}=1-P_{\bar{q}_1\bar{q}_3}-P_{\bar{q}_2\bar{q}_3}.
\end{aligned}
\end{eqnarray}
The summation index $\xi$ encompasses all isospin-spin intermediate configurations $\{I_{12},J_{12},I_3,J_3\}$
that can be coupled into the total quantum numbers of the state $H_{bbb}$. The term $\phi(\boldsymbol{\lambda})$
is the wave function for the relative motion between the state $T_{bb}$ and $\bar{B}^{*}$. The Jaccobi coordinate
$\boldsymbol{\lambda}$ can be explicitly expressed as
\begin{eqnarray}
\begin{aligned}
\boldsymbol{\lambda}=\frac{m_b\mathbf{r}_{b_1}+m_q\mathbf{r}_{\bar{q}_1}+m_b\mathbf{r}_{b_2}
+m_q\mathbf{r}_{\bar{q}_2}}{2(m_b+m_{\bar{q}})}
-\frac{m_b\mathbf{r}_{b_3}+m_{\bar{q}}\mathbf{r}_{{\bar{q}}_3}}{m_b+m_{\bar{q}}}.\nonumber
\end{aligned}
\end{eqnarray}

Accurate model calculations of the spatial components are essential for a precise
understanding of hadron properties. We employ the Gaussian expansion method (GEM),
which has proven to be highly effective for solving the few-body problem in nuclear
and hadron physics~\cite{Hiyama:2003cu,Hiyama:2018ivm}, to calculate the states
$\bar{B}\bar{B}^*\bar{B}^*$. In the GEM framework, the relative motion wave functions
$\phi(\mathbf{x})$ are expressed as a superposition of Gaussian functions with varying
sizes but identical angular excitation. This can be represented mathematically as follows,
\begin{eqnarray}
\phi(\mathbf{x})=\sum_{n=1}^{n_{max}}c_nN_{nl}x^{l}
e^{-\nu_nx^2}Y_{lm}(\hat{\mathbf{x}}),
\end{eqnarray}
where $\mathbf{x}$ encompasses the coordinates $\boldsymbol{r}$, $\boldsymbol{\rho}$,
and $\boldsymbol{\lambda}$. The associated orbital angular momentum for these coordinates
is denoted as $l_r$, $l_{\rho}$, and $l_{\lambda}$, respectively. More details regarding
the GEM can be found in Ref.~\cite{Hiyama:2003cu}.

\section{dimeson states $\boldsymbol{\bar{B}^{(*)}\bar{B}^* }$}

The initial step in studying exotic mesons involves accommodating ordinary mesons within
the quark model to determine the model parameters. By accurately solving the two-body Schr\"{o}dinger
equation, we have successfully reproduced the mass spectrum of ordinary mesons, allowing us to
establish the model parameters as detailed in Ref.~\cite{Deng:2014gqa}. The results for the
heavy-light mesons are summarized in Table~\ref{mesons}.
\begin{table} [h]
\caption{$Q\bar{q}$ meson spectrum, $\langle r^2\rangle^{\frac{1}{2}}$ is the size of mesons, mass
unit in MeV and $\langle r^2\rangle^{\frac{1}{2}}$ unit in fm.}
\label{mesons}
\tabcolsep=0.14cm
\begin{tabular}{cccccccccccccccccc}
\toprule[0.8pt]\noalign{\smallskip}
State&$D$&$D^*$&$D_s$&$D_s^*$&$\bar{B}$&$\bar{B}^*$&$\bar{B}_s$&$\bar{B}_s^*$\\
\toprule[0.8pt]\noalign{\smallskip}
Model&1867&2002&1972&2140&5259&5301&5377&5430\\
\noalign{\smallskip}
PDG&1869&2007&1968&2112&5280&5325&5366&5416\\
$\langle r^2\rangle^{\frac{1}{2}}$&0.68&0.82&0.52&0.69&0.73&0.77&0.57&0.62\\
\toprule[0.8pt]\noalign{\smallskip}
\end{tabular}
\end{table}

We have utilized the quark model to systematically investigate the doubly heavy tetraquark states,
focusing on both molecular and diquark configurations. Additionally, we have examined the correlation
between these two configurations~\cite{Deng:2018kly,Deng:2021gnb,Lin:2023gzm,Deng:2022cld}. The state
$T^+_{cc}$, observed by the LHCb Collaboration, can be described as a loosely bound deuteronlike state
within the quark model, which aligns remarkably well with the experimental data~\cite{Deng:2021gnb}.

The application of the quark model to the state $T^+_{cc}$ has demonstrated its feasibility to some extent.
In the following, we investigate the natures of the ground dimeson states $\bar{B}^{(*)}\bar{B}^*$
in the quark model, setting the stage for the investigations into the trimeson states. The dimeson
states have five possible isospin-spin configurations satisfying Einstein-Bose statistics, in which
$[\bar{B}\bar{B}^*]^1_0$ and $[\bar{B}^*\bar{B}^*]^1_0$ are isospin antisymmetric while
$[\bar{B}^*\bar{B}^*]^0_1$, $[\bar{B}\bar{B}^*]^1_1$, and $[\bar{B}^*\bar{B}^*]^2_1$
are isospin symmetric. To determine their eigenvalues and eigenvectors, we accurately solve
the four-body Schr\"{o}dinger equation
\begin{eqnarray}
(H_4-E_4)\Psi^{T_{bb}}_{I_{12}J_{12}}=0\nonumber
\end{eqnarray}
with the Rayleigh-Ritz variational principle.

Subsequently, we can calculate their binding energy,
\begin{eqnarray}
\Delta E_4=E_4-E_{\rm threshold},\nonumber
\end{eqnarray}
where $E_{\rm threshold}$ denotes the strong interaction threshold, which is the mass sum of
its constituents $\bar{B}^{(*)}$ and $\bar{B}^{*}$ in the quark model. If $\Delta E_4\geq0$,
the state is unbound and may dissociate into two mesons via the strong interactions. Conversely,
if $\Delta E_4<0$, the strong decay into two mesons is forbidden, leaving only the weak or
electromagnetic interactions as possible decay mechanisms.

To elucidate the binding mechanism of the bound state, we analyze the contributions from various
interactions and kinetic energy to $\Delta E_4$ using its eigenvector,
\begin{eqnarray}
\begin{aligned}
\Delta\langle V^{\chi}\rangle&=\langle\Psi^{T_{bb}}_{I_{12}J_{12}}|V^{\chi}|\Psi^{T_{bb}}_{I_{12}J_{12}}\rangle
-\langle\Phi^{\bar{B}^{(*)}}_{I_1J_1}|V^{\chi}|\Phi^{\bar{B}^{(*)}}_{I_1J_1}\rangle\\
&-\langle\Phi^{\bar{B}^{*}}_{I_2J_2}|V^{\chi}|\Phi^{\bar{B}^{*}}_{I_2J_2}\rangle,
\end{aligned}
\end{eqnarray}
where $\chi$ represents all types of interactions in the quark model. Additionally, to reveal
its spatial configuration, we calculate the size of the dimeson state $T_{bb}$. The numerical
results of these calculations are presented in Table~\ref{tbb}.

\begin{table*}[ht]
\caption{Binding energy $\Delta E_4$ of the dimeson states relative to their constituents
$\bar{B}^{(*)} \bar{B}^* $ and the contribution from various interactions and kinetic energy.
$\Delta V^{\rm con}$, $\Delta V^{\rm coul}$, $\Delta V^{\rm cm}$, $\Delta T$, $\Delta V^{\sigma}$,
$\Delta V^{\pi}$, and $\Delta V^{\eta}$ represent confinement term, Coulomb term, chromomagnetic
term, kinetic energy, $\sigma$-, $\pi$-, and $\eta$-meson exchange term, respectively, unit
in MeV. $\langle\boldsymbol{\rho}^2\rangle^{\frac{1}{2}}$ is the size of $T_{bb}$, unit in fm.}\label{tbb}
\tabcolsep=0.37cm
\begin{tabular}{ccccccccccccccccccccccc}
\toprule[0.8pt]\noalign{\smallskip}
~$T_{bb}$~&$I_{12}J_{12}^P$&$\Delta E_4$&$\Delta V^{\rm con}$&$\Delta V^{\rm coul}$&$\Delta V^{\rm cm}$&$~~\Delta T~~$
&$\Delta V^{\sigma}$&$\Delta V^{\pi}$&$\Delta V^{\eta}$&$\langle\boldsymbol{\rho}^2\rangle^{\frac{1}{2}}$\\
\noalign{\smallskip}\toprule[0.8pt]
\noalign{\smallskip}
\multirow{1}{*}{$[\bar{B}\bar{B}^*]^1_0$}
&$01^+$& $-10.0$ & $-6.3$ & $-8.9$ &  $-14.3$  &  33.0  & $-9.3$ & $-4.4$ &  $0.2$  & 1.07        \\
\noalign{\smallskip}
$[\bar{B}^*\bar{B}^*]^1_0$&$01^+$
& $-9.0$ & $-6.0$ & $-7.8$ & $-12.3$ & $29.4$ & $-8.5$ & $-3.9$  & $0.2$ & 1.11    \\
\noalign{\smallskip}
\toprule[0.8pt]
\end{tabular}
%\end{table*}
%\begin{table*}[ht]
\caption{Binding energy $\Delta E_6$ of the trimeson states relative to their constituents
$\bar{B}\bar{B}^*\bar{B}^*$ and the contribution from various interactions and kinetic energy.
$\langle\boldsymbol{\rho}^2\rangle^{\frac{1}{2}}$ and $\langle\boldsymbol{\lambda}^2\rangle^{\frac{1}{2}}$
represent the size of $T_{bb}$ and the distance between $\bar{B}^{(*)}$ and $T_{bb}$,
respectively, unit in fm.}\label{trimesons}
\tabcolsep=0.20cm
\begin{tabular}{ccccccccccccccccccccccc}
\toprule[0.8pt]\noalign{\smallskip}
~$T_{bb}$~&$H_{bbb}$&$IJ^P$&$\Delta E_6$&$\Delta V^{\rm con}$&$\Delta V^{\rm coul}$&$\Delta V^{\rm cm}$&$~~\Delta T~~$
&$\Delta V^{\sigma}$&$\Delta V^{\pi}$&$\Delta V^{\eta}$&$\langle\boldsymbol{\rho}^2\rangle^{\frac{1}{2}}$
&$\langle\boldsymbol{\lambda}^2\rangle^{\frac{1}{2}}$\\
\noalign{\smallskip}\toprule[0.8pt]
\noalign{\smallskip}
\multirow{5}{*}{$[\bar{B}\bar{B}^*]^1_0$}& $\left[[\bar{B}\bar{B}^*]^1_0\bar{B}^*\right]^0_{\frac{1}{2}}$&$\frac{1}{2}0^-$
& $-10.2$  &  $-6.6$  &  $-9.2$  &  $-14.0$  &  34.6 &  $-10.8$  &  $-4.4$ &   0.3   & 1.09   & $4.75$\\
\noalign{\smallskip}
& $\left[[\bar{B}\bar{B}^*]^1_0\bar{B}^*\right]^1_{\frac{1}{2}}$&$\frac{1}{2}1^-$
& $-10.0$ & $-6.3$ & $-8.9$ &  $-14.3$  &  33.0  & $-9.3$ & $-4.4$ &  $0.2$  & 1.07 & $\infty$ \\
\noalign{\smallskip}
& $\left[[\bar{B}\bar{B}^*]^1_0\bar{B}^*\right]^2_{\frac{1}{2}}$&$\frac{1}{2}2^-$
& $-10.0$ & $-6.3$ & $-8.9$ &  $-14.3$  &  33.0  & $-9.3$ & $-4.4$ &  $0.2$  & 1.07 & $\infty$\\
\noalign{\smallskip}
\noalign{\smallskip}

\multirow{4}{*}{$[\bar{B}\bar{B}^*]^1_1$}
& $\left[[\bar{B}\bar{B}^*]^1_1\bar{B}^*\right]^0_{\frac{1}{2}}$&$\frac{1}{2}0^-$
& $-0.4$  &  $-2.0$  &  $-3.1$  &  $-1.1$  &  15.2  &  $-10.2$  &  $0.3$ &   0.5   & 2.19   & $1.49$\\
\noalign{\smallskip}
& $\left[[\bar{B}\bar{B}^*]^1_1\bar{B}^*\right]^1_{\frac{1}{2}}$&$\frac{1}{2}1^-$
& $-0.6$ & $-3.7$ & $-4.9$ & $-6.7$ & $24.2$ & $-8.2$ & $-1.5$  & $0.1$& $2.64$ & $1.28$  \\
\noalign{\smallskip}
& $\left[[\bar{B}\bar{B}^*]^1_1\bar{B}^*\right]^2_{\frac{1}{2}}$&$\frac{1}{2}2^-$
&unbound&&&&&&&&$\infty$&$\infty$       \\
\noalign{\smallskip}
\noalign{\smallskip}

$[\bar{B}^*\bar{B}^*]^1_0$
& $\left[\bar{B} [\bar{B}^*\bar{B}^*]^1_0\right]^1_{\frac{1}{2}}$&$\frac{1}{2}1^-$
& $-9.0$ & $-6.0$ & $-7.8$ & $-12.3$ & $29.4$ & $-8.5$ & $-3.9$  & $0.2$ & 1.11& $\infty$  \\
\noalign{\smallskip}
\noalign{\smallskip}

\multirow{1}{*}{$[\bar{B}^*\bar{B}^*]^0_1$}
& $\left[\bar{B} [\bar{B}^*\bar{B}^*]^0_1\right]^0_{\frac{1}{2}}$&$\frac{1}{2}0^-$
& $-0.7$  &  $-3.1$  &  $-2.2$  &  $-4.4$  &  22.5 &  $-11.7$  &  $-1.9$ &   0.2   & 2.09   & $1.43$\\
\noalign{\smallskip}
\noalign{\smallskip}

$[\bar{B}^*\bar{B}^*]^2_1$
& $\left[\bar{B} [\bar{B}^*\bar{B}^*]^2_1\right]^2_{\frac{1}{2}}$&$\frac{1}{2}2^-$
&unbound&&&&&&&&$\infty$&$\infty$   \\
\noalign{\smallskip}
\toprule[0.8pt]
\end{tabular}
\end{table*}

The energy of the configuration $[\bar{B}\bar{B}^*]_0^1$ is approximately 10.0 MeV lower
than its lowest threshold $\bar{B}\bar{B}^*$, indicating that this configuration is a stable
state against strong interactions, as detailed in Table~\ref{tbb}. Conversely, the energy
of the configuration $[\bar{B}^*\bar{B}^*]_0^1$ is about 9.0 MeV below the $\bar{B}^*\bar{B}^*$
threshold, yet it remains above the lowest threshold $\bar{B}\bar{B}^*$. Consequently, the
$[\bar{B}^*\bar{B}^*]_0^1$ configuration is better characterized as a resonance rather
than a bound state within the quark model, owing to its potential decay into $\bar{B}$
and $\bar{B}^*$ through strong interactions.

In these configurations, the attraction arises from the interactions denoted as $V^{\sigma}$,
$V^{\pi}$, $V^{\rm con}$, $V^{\rm cm}$ and $V^{\rm coul}$ within the quark model.
The relative distances between $\bar{B}^{(*)} $ and $\bar{B}^*$ in the states
$[\bar{B}^{(*)}\bar{B}^*]_0^1$ are approximately 1.10 fm, while the typical sizes of
the mesons $\bar{B}$ and $\bar{B}^*$ range between 0.70 and 0.80 fm, as shown in Tables~\ref{mesons}
and\ref{tbb}. The significant overlap between $\bar{B}^{(*)}$ and $\bar{B}^*$ suggests
that the configuration $[\bar{B}^{(*)}\bar{B}^*]_0^1$ resembles a compact tetraquark state.
This conclusion is supported by the other model studies of the state $T^-_{bb}$ with $01^+$
and such meson-meson configuration~\cite{Yang:2019itm,Meng:2020knc,Chen:2021tnn,Ortega:2024epk}.

In general, the meson-meson configuration is expected to form a loosely bound
molecule through residual interactions between the two colorless subclusters. The model
studies of the meson-meson configuration $[c\bar{q}][c\bar{q}]$ indeed indicated its
loose molecular structure~\cite{Yang:2019itm,Meng:2020knc,Chen:2021tnn,Ortega:2024epk},
which is in strong agreement with the deuteronlike state $T_{cc}^+$ reported by the LHCb.
In stark contrast, the meson-meson configuration $[b\bar{q}][b\bar{q}]$ forms
a compact tetraquark structure within the same
framework~\cite{Yang:2019itm,Meng:2020knc,Chen:2021tnn,Deng:2021gnb,Deng:2022cld,Ortega:2024epk}.
The primary reason for this is that the large mass of the $b$-quarks helps to
reduce the system's kinetic energy, allowing the two subclusters to come closer
together. This proximity enables the model to provide stronger attractions than in
the meson-meson configuration $[c\bar{q}][c\bar{q}]$. These observations suggest that
the final structure of the systems is not only influenced by the initial assumptions
or trial wave function, but also by the intrinsic properties of the system
and the dynamics inherent in the model.

The three isospin symmetric configurations, $[\bar{B}^*\bar{B}^*]^0_1$, $[\bar{B}\bar{B}^*]^1_1$,
and $[\bar{B}^*\bar{B}^*]^2_1$, are unable to form stable bound states due to the absence of
binding forces in the quark model. Specifically, the interactions between $\bar{B}^{(*)}$ and
$\bar{B}^*$ are repulsive when these particles are arranged as an isospin triplet in the finite
space. Lattice QCD studies focusing on these isospin symmetric configurations have shown no
detectable signals of states existing below their respective thresholds, indicating a lack
of binding forces~\cite{Junnarkar:2018twb}. Generally, it is found that all doubly heavy tetraquark
states, whether isospin, $V$-spin, or $U$-spin symmetric, do not result in stable states
due to the strong interactions~\cite{Deng:2021gnb,Junnarkar:2018twb,Li:2012ss}. Contrasting
these findings, however, the isospin symmetric state $T_{bb}$ does exhibit a binding
energy on the order of several MeVs in the models using the one-pion exchange potential~\cite{Ma:2018vhp}.

\section{trimeson states $\boldsymbol{\bar{B}\bar{B}^*\bar{B}^*}$}

We extend the model calculations to the trimeson system $\bar{B}\bar{B}^*\bar{B}^*$
to search for potential bound states. Generally, orbitally excited bound states are more
challenging to form due to the repulsive centrifugal potential. Therefore, this study will
concentrate on the ground states of the trimeson system.

\begin{figure} [h]
\centering
\resizebox{0.49\textwidth}{!}{\includegraphics{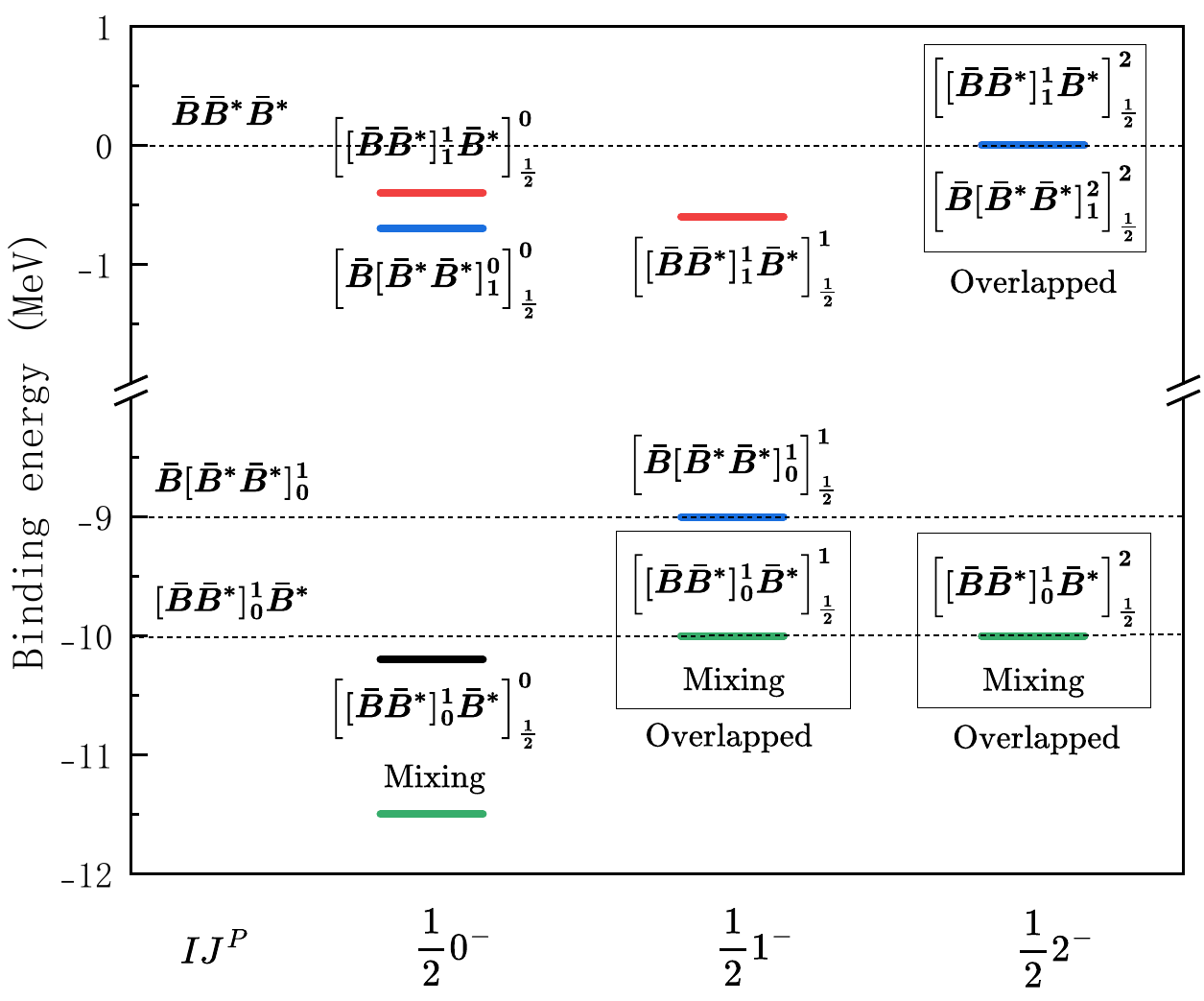}}
\caption{The binding energy spectrum of the trimeson states, relative to their constituent
mesons $\bar{B}\bar{B}^*\bar{B}^*$, is presented. The thresholds, $\bar{B}\bar{B}^*\bar{B}^*$,
$[\bar{B}\bar{B}^*]_0^1\bar{B}^*$, and $\bar{B}[\bar{B}^*\bar{B}^*]_0^1$
are marked by three dotted horizontal lines.}
\label{spectrum}
\end{figure}

Based on the dimeson states $\bar{B}^{(*)}\bar{B}^*$, we construct all possible trimeson
configurations with $I=\frac{1}{2}$, which are summarized in Table~\ref{trimesons}.
We omit the trimeson states with $I=\frac{3}{2}$ since they cannot form any bound states
due to the absence of attractive forces in the dimeson configurations with $I=1$ in our model.
To obtain the eigenvalues and eigenvectors of the trimeson configurations, we solve the
six-body Schr\"{o}dinger equations for the bound-state problem. Following the same
procedures used for the dimeson states, we calculate their binding energy relative
to their constituent mesons $\bar{B}\bar{B}^*\bar{B}^*$, along with the contributions
to the binding energy from various components of the Hamiltonian. We also examine the
size of the subcluster $T_{bb}$  and the distance between the subclusters $\bar{B}^{(*)}$
and $T_{bb}$. Those numerical results are presented in Table~\ref{trimesons}. To provide a more intuitive
representation of the energy spectrum, the binding energy levels of the trimeson
states with various isospin-spin configurations, relative to their constituent
mesons $\bar{B}\bar{B}^*\bar{B}^*$, are displayed in Fig.~\ref{spectrum}.

\subsection{$\boldsymbol{H_{bbb}}$ from $\boldsymbol{[\bar{B}\bar{B}^*]^1_0}$ and $\boldsymbol{[\bar{B}\bar{B}^*]^1_1}$}

From Table~\ref{trimesons} and FIG.~\ref{spectrum}, it can be observed that the binding
energy of the isospin-spin configuration $[[\bar{B} \bar{B}^*]^1_0\bar{B}^*]^0_{\frac{1}{2}}$
is approximately $-10.2$ MeV relative to its constituent $\bar{B}\bar{B}^*\bar{B}^* $.
The majority of this binding energy originates from the subcluster $[\bar{B} \bar{B}^*]^1_0$,
which contributes about $-10.0$ MeV. In contrast, the remaining contribution,
which is approximately $-0.2$ MeV, indicates a weak attraction between the
subclusters $[\bar{B}\bar{B}^*]_0^1$ and $\bar{B}$. As a result, the configuration
is stable against the strong interactions; it cannot directly decay into its constituent
$\bar{B}\bar{B}^*\bar{B}^*$ nor into $[\bar{B}\bar{B}^*]_0^1$ and $\bar{B}^*$. However,
the configuration is about 80 MeV above the trimeson threshold $\bar{B}\bar{B}\bar{B}$,
allowing it to decay into $\bar{B}\bar{B}\bar{B}$ through electromagnetic
interactions. Furthermore, it is approximately 400 MeV above the $\Omega_{bbb}$-$\bar{p}$
threshold according to model predictions, enabling it to decay into $D$-wave
$\Omega_{bbb}$-$\bar{p}$ states via the strong interactions. These decay
properties of the configuration deserves further investigation in the future.

\begin{figure} [h]
\centering
\resizebox{0.45\textwidth}{!}{\includegraphics{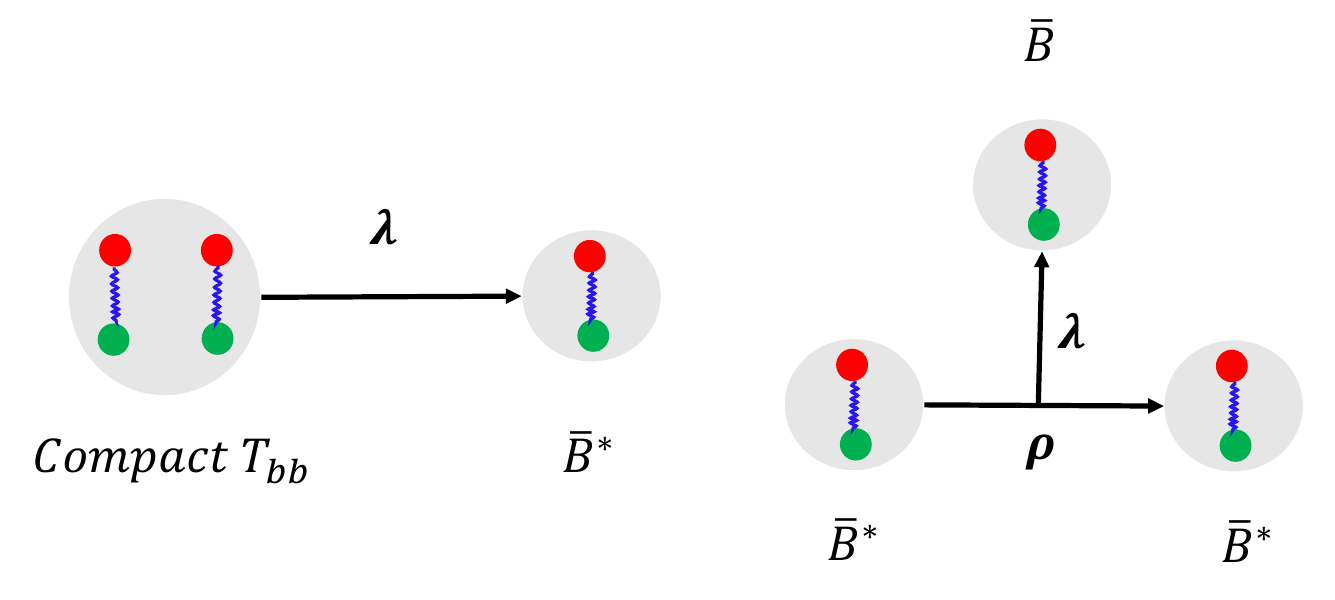}}
\caption{Spatial configurations of the trimeson state $\bar{B}\bar{B}^{*}\bar{B}^* $,
two-body molecule (left) and three-body molecule (right). A red ball represents a $b$-quark,
while a green ball represents either a $\bar{u}$-quark or a $\bar{d}$-quark.}
\label{hbbb}
\end{figure}

The average size of the subcluster $[\bar{B}\bar{B}^*]^1_0$ within the
configuration $[[\bar{B}\bar{B}^*]^1_0\bar{B}^*]^0_{\frac{1}{2}}$ is
approximately 1.09 fm, which is nearly identical to that of the isolated
$[\bar{B}\bar{B}^*]^1_0$, see $\langle\boldsymbol{\rho}^2\rangle^{\frac{1}{2}}$
in Tables~\ref{tbb} and\ref{trimesons}. This observation suggests
that the characteristics of $[\bar{B}\bar{B}^*]^1_0$ as a building
block remain largely unchanged. The distance between the two subclusters
$[\bar{B}\bar{B}^*]_0^1$ and $\bar{B}^*$ is about 4.75 fm, which is significantly
larger than their sizes. Consequently, the configuration can be described
as a loose two-body molecular state composed of a compact $[\bar{B}\bar{B}^*]^1_0$
and $\bar{B}^*$, rather than an evident three-body configuration, see
Fig.~\ref{hbbb}. By comparing the binding energies of $[\bar{B}\bar{B}^*]^1_0$
and $[[\bar{B}\bar{B}^*]^1_0\bar{B}^*]^0_{\frac{1}{2}}$, along
with the contributions from each component, it can be concluded that
the binding mechanism of the configuration primarily arises from the
$\sigma$ meson exchange between the subclusters $[\bar{B}\bar{B}^*]_0^1$
and $\bar{B}^*$.

Comparing the configuration $\left[[\bar{B}\bar{B}^*]^1_0\bar{B}^*\right]^1_{\frac{1}{2}}$
with the configuration $[\bar{B}\bar{B}^*]^1_0$, we find that they share exactly
the same binding energy and contributions from various interactions in the model.
This suggests that the inclusion of $\bar{B}^* $ does not provide any additional
attraction. Furthermore, the distance between the subclusters $[\bar{B}\bar{B}^*]^1_0$
and $\bar{B}^*$ approaches infinity. These observations clearly indicate that
the meson $\bar{B}^*$, as a spectator, cannot participate in the configuration
$[\bar{B}\bar{B}^*]^1_0$ to form a bound trimeson configuration
$[[\bar{B}\bar{B}^*]^1_0\bar{B}^*]^1_{\frac{1}{2}}$ in the quark model.
For the same reasons, the trimeson configuration
$[[\bar{B} \bar{B}^*]^1_0\bar{B}^*]^2_{\frac{1}{2}}$ also cannot establish
a trimeson bound state in the model.

The energy of the configuration $[[\bar{B} \bar{B}^*]^1_1\bar{B}^*]^0_{\frac{1}{2}}$
is about 0.4 MeV lower than that of its constituent $\bar{B}\bar{B}^*\bar{B}^*$,
but 9.6 MeV higher than the combined energy of the configuration $[\bar{B}\bar{B}^*]^1_0$
and $\bar{B}^*$. This indicates that it can form a stable trimeson state against decay
into its constituents within the model. Since $\bar{B} $ and $\bar{B}^*$ in the state
$[\bar{B} \bar{B}^*]^1_1$ exhibit repulsive interactions in the finite space, we can
reasonably infer that the binding forces in the configuration
$[[\bar{B} \bar{B}^*]^1_1\bar{B}^*]^0_{\frac{1}{2}}$ arise from the attractive
interactions between the subclusters $[\bar{B}\bar{B}^*]^1_1$ and $\bar{B}^*$.
The $\sigma$ meson exchange interaction remains a significant contributor to this attraction.

The size of the subcluster $[\bar{B}\bar{B}^*]^1_1$ is approximately 2.19 fm,
while the distance between $[\bar{B}\bar{B}^*]^0_1$ and $\bar{B}^*$ is around
1.49 fm (refer to $\langle\boldsymbol{\rho}^2\rangle^{\frac{1}{2}}$ and
$\langle\boldsymbol{\lambda}^2\rangle^{\frac{1}{2}}$ in Table~\ref{trimesons}).
These sizes are evidently larger than that of the meson $\bar{B}^{(*)}$. The three mesons
are completely separated, resulting in a triangular spatial structure for the state.
Thus, the configuration $[[\bar{B}\bar{B}^*]^1_1\bar{B}^*]^0_{\frac{1}{2}}$
can be characterized as a loose three-body molecular state, as shown in Fig.~\ref{hbbb}.

The energy of the configuration $[[\bar{B}\bar{B}^*]^1_1\bar{B}^*]^1_{\frac{1}{2}}$
is about 0.6 MeV lower than that of its constituent $\bar{B}\bar{B}^*\bar{B}^*$. Its
spatial configuration and properties are similar to those of the configuration
$[[\bar{B}\bar{B}^*]^1_1\bar{B}^*]^0_{\frac{1}{2}}$. In contrast, the configuration
$[[\bar{B} \bar{B}^*]^1_1\bar{B}^*]^2_{\frac{1}{2}}$ is unbound within the model
because both the size of the subcluster $[\bar{B}\bar{B}^*]_1^1$ and the distance
between the subclusters $[\bar{B}\bar{B}^*]_1^1$ and $\bar{B}^*$ tend to infinity,
as detailed in Table~\ref{trimesons} and FIG.~\ref{spectrum}.

\subsection{$\boldsymbol{H_{bbb}}$ from $\boldsymbol{[\bar{B}^*\bar{B}^*]^1_0}$,
$\boldsymbol{[\bar{B}^*\bar{B}^*]^0_1}$, and $\boldsymbol{[\bar{B}^*\bar{B}^*]^2_1}$}

The configuration $[\bar{B}^*\bar{B}^*]^1_0$ in Table~\ref{tbb} and the configuration
$\left[\bar{B}[\bar{B}^*\bar{B}^*]^1_0\right]^1_{\frac{1}{2}}$ in Table~\ref{trimesons}
exhibit precisely identical binding energies, as well as equivalent contributions from various
interactions and kinetic energy. Moreover, the distance between the subclusters
$[\bar{B}^*\bar{B}^*]^1_0$ and $\bar{B}$ approaches infinity. As a result, the
subclusters $[\bar{B}^*\bar{B}^*]^1_0$ and $\bar{B}$ cannot form a bound trimeson
state within the model. Additionally, the configuration $[\bar{B}^*\bar{B}^*]^2_1$
and the meson $\bar{B} $ also do not lead to a stable configuration
$\left[\bar{B}[\bar{B}^*\bar{B}^*]^2_1\right]^2_{\frac{1}{2}}$ in the model, as shown
in FIG.~\ref{spectrum}.

The energy of the configuration $[\bar{B}[\bar{B}^*\bar{B}^*]^0_1]^0_{\frac{1}{2}}$ is
approximately 0.7 MeV lower than that of its constituent $\bar{B}\bar{B}^*\bar{B}^*$,
allowing for the establishment of a stable trimeson state that resists disintegration into its
constituents within the model. Since $\bar{B}^*$ and $\bar{B}^*$ in the configuration
$[\bar{B}^*\bar{B}^*]^0_1$ are mutually repulsive in the finite space, it is reasonable
to infer that the binding forces in the configuration
$[\bar{B}[\bar{B}^*\bar{B}^*]^0_1]^0_{\frac{1}{2}}$ arise from the attractive
interactions between the subclusters $\bar{B}$ and $[\bar{B}^*\bar{B}^*]^0_3$.
More specifically, the binding forces originate from the attractions between $\bar{B}$
and $\bar{B}^*$. In this context, $\bar{B}$ acts as a core that effectively binds the
repulsive $\bar{B}^*$ and $\bar{B}^*$ together, resulting in the formation of a trimeson
bound state, where the exchange interaction mediated by the $\sigma$ meson plays a pivotal role.

The size of the subcluster $[\bar{B}^*\bar{B}^*]^0_1$ within this configuration
$[\bar{B}[\bar{B}^*\bar{B}^*]^0_1]^0_{\frac{1}{2}}$ is approximately 2.09 fm,
while the distance between the subclusters $\bar{B} $ and $[\bar{B}^*\bar{B}^*]^0_1$
is about 1.47 fm, as indicated by $\langle\boldsymbol{\rho}^2\rangle^{\frac{1}{2}}$ and
$\langle\boldsymbol{\lambda}^2\rangle^{\frac{1}{2}}$ in Table~\ref{trimesons}.
Thus, the configuration $[\bar{B} [\bar{B}^*\bar{B}^*]^0_1]^0_{\frac{1}{2}}$
represents a loosely three-body molecular state. Similar to the configuration
$\left[[\bar{B}\bar{B}^*]^1_0\bar{B}^*\right]^0_{\frac{1}{2}}$, the configuration
$[\bar{B}[\bar{B}^*\bar{B}^*]^0_1]^0_{\frac{1}{2}}$ can decay into the states
$\bar{B}\bar{B}\bar{B}$ and $D$-wave $\Omega_{bbb}$-$\bar{p}$. Additionally, this
state may also decay into the $S$-wave dimeson state $[\bar{B}\bar{B}^*]^1_0$ and
$\bar{B}^*$ through rearrangements of its constituents driven by strong interactions.

\subsection{Coupled channel effects}

From a quantum mechanical perspective, the trimeson states should encompass linear
combinations of all possible isospin-spin configurations allowed by its quantum numbers.
The mixing of these configurations possibly depress the energy of the states. In this
context, we perform coupling calculations involving three configurations that have
the same isospin and spin quantum numbers. After coupling isospin-spin configurations
with $\frac{1}{2}0^-$, the final energy of the trimeson state with $\frac{1}{2}0^-$
is approximately $-11.5$ MeV below its constituent mesons $\bar{B}\bar{B}^*\bar{B}^*$,
as shown in FIG.~\ref{spectrum}. In other words, the coupled channel effect enhances
the attraction between two subclusters, $\bar{B}$ and $[\bar{B}\bar{B}^*]_0^1$
in the trimeson state with $0\frac{1}{2}^-$, by about 1.3 MeV. Despite this, the
trimeson state with $\frac{1}{2}0^-$ remains a loosely two-body bound state, with
the distance between the two subclusters being approximately 2.2 fm. The probabilities
of the isospin-spin configurations $\left[[\bar{B}\bar{B}^*]^1_0\bar{B}^*\right]^0_{\frac{1}{2}}$
and $\left[\bar{B}[\bar{B}^*\bar{B}^*]^0_1\right]^0_{\frac{1}{2}}$ in the trimeson
state with $0\frac{1}{2}^-$ are approximately $80\%$ and $14\%$ , respectively. In
contrast, the configuration $\left[[\bar{B}\bar{B}^*]^1_1\bar{B}^*\right]^0_{\frac{1}{2}}$
contributes only $6\%$. In the case of the trimeson states with $\frac{1}{2}1^-$ and
$\frac{1}{2}2^-$, the coupling between the three isospin-spin configurations is insufficient
to lower their lowest energy.

\subsection{Correlations of meson pairs}

The trimeson ground states $\bar{B}\bar{B}^*\bar{B}^*$ with $IJ^P=\frac{1}{2}0^-$
can exist in three different isospin-spin configurations:
$\left[[\bar{B}\bar{B}^*]^1_0\bar{B}^*\right]^0_{\frac{1}{2}}$,
$\left[[\bar{B}\bar{B}^*]^1_1\bar{B}^*\right]^0_{\frac{1}{2}}$, and
$\left[\bar{B} [\bar{B}^*\bar{B}^*]^0_1\right]^0_{\frac{1}{2}}$. All of these configurations
can establish stable trimeson states that resist disintegration into their constituent particles
within the model. Notably, the energy of the configuration
$\left[[\bar{B}\bar{B}^*]^1_0\bar{B}^*\right]^0_{\frac{1}{2}}$ is clearly lower than
that of other two configurations.

It is important to emphasize that the isospin-spin configuration
$\left[[\bar{B} \bar{B}^*]^1_0\bar{B}^*\right]^0_{\frac{1}{2}}$ represents a loose
two-body molecular state rather than a loose three-body state, where the mesons $\bar{B} $
and $\bar{B}^*$ can form a compact tetraquark state denoted as $[\bar{B}\bar{B}^*]^1_0$.
More precisely, this configuration is a bound state composed of the compact tetraquark state
$[\bar{B} \bar{B}^*]^1_0$  and the meson $\bar{B}^*$. In contrast, the other two
isospin-spin configurations are characterized as loose three-body molecular states, formed
by the completely separated mesons. Those evidences clearly indicate that the strong
correlation of the meson pair $[\bar{B}\bar{B}^*]^1_0$ prevails over those of the pairs
$[\bar{B}\bar{B}^*]^1_1$ and $[\bar{B}^*\bar{B}^*]^0_1$ in the formation of the trimeson
bound state $\bar{B}\bar{B}^*\bar{B}^*$ with $\frac{1}{2}0^-$.

The correlation of the meson pair closely resembles the nucleon-nucleon correlations
observed in the three-nucleon system of $^3\text{He}$, where the $p$-$n$ pairs are predominant,
constituting approximately eighty percent of the correlations~\cite{Li:2022fhh}.
In other words, the probability of forming $p$-$n$ pairs is four times that of $p$-$p$
pairs. This proportion is significantly higher than that of $p$-$p$ pairs. The $80\%$
probability of the isospin-spin configuration $\left[[\bar{B}\bar{B}^*]^1_0\bar{B}^*\right]^0_{\frac{1}{2}}$
in the trimeson state with $\frac{1}{2} 0^-$ suggests that the strongly correlated
meson pair $[\bar{B}\bar{B}^*]^1_0$ predominates over other possible meson pair configurations.

To unambiguously analyze the binding mechanism between the subclusters $[\bar{B}\bar{B}^*]^1_0$
and $\bar{B}^*$ in the isospin-spin configuration $\left[[\bar{B}\bar{B}^*]^1_0\bar{B}^*\right]^0_{\frac{1}{2}}$
from the correlation of the meson pairs, we assign numbers to three constituent particles
and rearrange the configuration $\left[[\bar{B}_1\bar{B}^*_2]^1_0\bar{B}^*_3\right]^0_{\frac{1}{2}}$
as follow,
\begin{eqnarray}
\begin{aligned}
\left[[\bar{B}_1\bar{B}^*_2]^1_0\bar{B}^*_3\right]^0_{\frac{1}{2}}
=\left[[\bar{B}_1\bar{B}^*_3]^1_0\bar{B}^*_2\right]^0_{\frac{1}{2}}
+\left[\bar{B}_1[\bar{B}^*_2\bar{B}^*_3]^{0,2}_0\right]^0_{\frac{1}{2}},\nonumber\\
\end{aligned}
\end{eqnarray}
It can be observed that the configuration includes another strongly correlated meson pair
$[\bar{B}_1\bar{B}^*_3]^1_0$, which is responsible for the binding force between the the
subclusters $[\bar{B}_1\bar{B}^*_2]^1_0$ and $\bar{B}^*_3$. It is important to note that
the orbital components of the strongly correlated meson pair $[\bar{B}_1\bar{B}^*_3]^1_0$
encompasses not only the ground state but also angular excitations, as determined by the
complex angular momentum algebra~\cite{Hiyama:2003cu,Deng:2022cld}. Furthermore, the meson pairs
$[\bar{B}^*_2\bar{B}^*_3]^{0,2}_0$ must be in the excitation because of both the Bose-Einstein
statistics and total quantum numbers. These angular excitations lead to an expansion of the
distance between the subclusters $[\bar{B}_1\bar{B}^*_2]^1_0$ and $\bar{B}^*_3$. Consequently,
the configuration $\left[[\bar{B}_1\bar{B}^*_2]^1_0\bar{B}^*_3\right]^0_{\frac{1}{2}}$
can only form a loose two-body bound state of the compact tetraquark state
$[\bar{B}_1\bar{B}^*_2]^1_0$ and $\bar{B}^*_3$.

We rearrange other two isospin-spin configurations
$\left[[\bar{B}_1\bar{B}^*_2]^1_1\bar{B}^*_3\right]^0_{\frac{1}{2}}$
and $\left[\bar{B}_1[\bar{B}^*_2\bar{B}^*_3]^0_1\right]^0_{\frac{1}{2}}$ as follow,
\begin{eqnarray}
\begin{aligned}
\left[[\bar{B}_1\bar{B}^*_2]^1_1\bar{B}^*_3\right]^0_{\frac{1}{2}}
&=\frac{1}{\sqrt{3}}\left(\left[[\bar{B}_1\bar{B}^*_3]^1_0\bar{B}^*_2\right]^0_{\frac{1}{2}}
+\left[\bar{B}_1[\bar{B}^*_2\bar{B}^*_3]^{0,2}_0\right]^0_{\frac{1}{2}}\right),\nonumber\\
\left[\bar{B}_1[\bar{B}^*_2\bar{B}^*_3]^0_1\right]^0_{\frac{1}{2}}
&=\frac{1}{\sqrt{3}}\left(\left[[\bar{B}_1\bar{B}^*_2]^1_0\bar{B}^*_3\right]^0_{\frac{1}{2}}
+\left[[\bar{B}_1\bar{B}^*_3]^1_0\bar{B}^*_2\right]^0_{\frac{1}{2}}\right),\nonumber\\
\end{aligned}
\end{eqnarray}
where the configuration $\left[\bar{B}_1[\bar{B}^*_2\bar{B}^*_3]^0_1\right]^0_{\frac{1}{2}}$
incorporates two strongly correlated meson pairs $[\bar{B}_1\bar{B}^*_2]^1_0$ and $[\bar{B}_1\bar{B}^*_3]^1_0$,
while the configuration $\left[[\bar{B}_1\bar{B}^*_2]^1_1\bar{B}^*_3\right]^0_{\frac{1}{2}}$
just involves only one strongly correlated meson pair $[\bar{B}_1\bar{B}^*_3]^1_0$. As a result,
the configuration $\left[\bar{B}_1[\bar{B}^*_2\bar{B}^*_3]^0_1\right]^0_{\frac{1}{2}}$ has a
lower binding energy, as illustrated in Table~\ref{trimesons} and FIG.~\ref{spectrum}. The
orbital components of those strongly correlated meson pairs in the two configurations encompass
not only the ground state but also include angular excitations. These angular excitations
lead to an expansion of the strongly correlated meson pairs, thereby weakening their attractive
interactions. As a result, these two configurations can only form a very shallow three-body
bound state.

Similarly, we can also rearrange the isospin-spin configuration
$\left[[\bar{B}_1\bar{B}^*_2]^1_1\bar{B}^*_3\right]^1_{\frac{1}{2}}$ as
\begin{eqnarray}
\begin{aligned}
\left[[\bar{B}_1\bar{B}^*_2]^1_1\bar{B}^*_3\right]^1_{\frac{1}{2}}
=\frac{1}{\sqrt{3}}\left(\left[[\bar{B}_1\bar{B}^*_2]^1_0\bar{B}^*_3\right]^1_{\frac{1}{2}}
+\left[\bar{B}_1[\bar{B}^*_2\bar{B}^*_3]^1_0\right]^1_{\frac{1}{2}}\right).\nonumber
\end{aligned}
\end{eqnarray}
This rearrangement incorporates one strongly correlated meson pairs $[\bar{B}_1\bar{B}^*_2]^1_0$ as
well as another attractive meson pair $[\bar{B}^*_2\bar{B}^*_3]^1_0$. It is important to emphasize that
the strongly correlated meson pair $[\bar{B}_1\bar{B}^*_2]^1_0$, rather the pair $[\bar{B}^*_2\bar{B}^*_3]^1_0$,
plays a decisive role in the stable configuration $\left[[\bar{B}_1\bar{B}^*_2]^1_1\bar{B}^*_3\right]^1_{\frac{1}{2}}$.
This is because the isospin-spin configuration $\left[\bar{B}_1[\bar{B}^*_2\bar{B}^*_3]^1_0\right]^1_{\frac{1}{2}}$
cannot establish a stable trimeson state, as detailed in Table~\ref{trimesons} and FIG.~\ref{spectrum}.

The following points need to be clarified regarding the above arrangement of the isospin-spin
wave functions in this work. For the trimeson state $\bar{B}_1\bar{B}_2^*\bar{B}_3^*$, only
the isospin part needs to be rearranged, while the spin part remains unchanged since the spin
of the meson $\bar{B}_1$ is zero. The isospin-spin wave functions are not expanded according
to some specific coupling order, where the attention should be paid to the particle numbers.
The expansions considered here differ from the transformation between two sets of wave functions
with specific coupling orders using the 6-j symbols. Therefore, the expansion coefficients are
not normalized in this work. \\

\section{summary and outlook}

In this study, we have investigated the properties of the trimeson states $\bar{B}\bar{B}^*\bar{B}^*$
with various isospin-spin configurations in the quark model utilizing the Gaussian expansion
method. By precisely solving the six-body Schr\"{o}dinger equations for the bound state
question, we put forward four possible stable isospin-spin configurations,
$\left[[\bar{B}\bar{B}^*]^1_0\bar{B}^*\right]^0_{\frac{1}{2}}$,
$\left[[\bar{B}\bar{B}^*]^1_1\bar{B}^*\right]^0_{\frac{1}{2}}$,
$\left[\bar{B} [\bar{B}^*\bar{B}^*]^0_1\right]^0_{\frac{1}{2}}$, and
$\left[[\bar{B}\bar{B}^*]^1_1\bar{B}^*\right]^1_{\frac{1}{2}}$, against dissociation into their
constituent particles $\bar{B}\bar{B}^ *\bar{B}^* $. Additionally, we analyze
their spatial configurations, underlying binding mechanisms and the correlations of meson pairs.

The configuration $\left[[\bar{B}\bar{B}^*]^1_0\bar{B}^*\right]^0_{\frac{1}{2}}$ is approximately
10.2 MeV lower than the energy of its constituent particles. Furthermore, this configuration is about
0.2 MeV below the threshold of the compact tetraquark state $[\bar{B}\bar{B}^*]^1_0$ and $\bar{B}^* $.
This configuration represents a loose two-body bound state composed of the subclusters $[\bar{B}\bar{B}^*]^1_0$
and $\bar{B}^* $, with a size of around 4.75 fm. In contrast, other three configurations exhibit binding
energies of less than 1 MeV relative to their constituent particles, forming a loose three-meson bound
state. The $\sigma$ meson exchange interaction acts as a decisive role in the formation of those four
stable configurations.

After coupling three isospin-spin configurations with $\frac{1}{2}0^-$, the final energy of
the trimeson state with $\frac{1}{2}0^-$ is approximately $-11.5$ MeV below its constituent
mesons. The trimeson state with $\frac{1}{2}0^-$ remains a loosely two-body bound state
$[\bar{B}\bar{B}^*]^1_0$ $\bar{B}^*$ with a binding energy around 1.5 MeV and a huge size of
2.20 fm. The main component of is the isospin-spin configuration
$\left[[\bar{B}\bar{B}^*]^1_0\bar{B}^*\right]^0_{\frac{1}{2}}$ in the trimeson
state and its probability is approximately $80\%$. In the case of the trimeson states
with $\frac{1}{2}1^-$ and $\frac{1}{2}2^-$, however, the coupling calculation can not push
down their energy.

The size of the meson pair $[\bar{B}\bar{B}^*]^1_0$ is obvious smaller than other types of meson
pairs in the four stable isospin-spin configurations. There exists a strong attractive interaction
in this meson pair. Such a strongly correlated meson pair resembles the short-range strong correlation
found in the $p$-$n$ pairs in nuclear physics. These short-range strong correlations are crucial for
a comprehensive understanding of nuclear dynamics and the forces at play at short distances, as well
as how they arise from the strong interactions among quarks within nucleons. Similarly, the strongly
correlated meson pair $[\bar{B}\bar{B}^*]^1_0$ plays an indispensable role in the formation of the
four stable isospin-spin configurations. They are not only responsible for the binding mechanisms
but also influences the spatial structures of those stable trimeson configurations at the meson
level.

The hexaquark system possesses a rich variety of color configurations, including the colorless
singlets and hidden color configurations, some of which do not exist in ordinary hadrons. The
physical effects of these color configurations, particularly the hidden color components, are
intriguing. To fully understand these effects, further investigation is required.

\acknowledgments{This research is supported by the Fundamental Research Funds for the Central Universities
under Grant No. SWU-XDJH202304.}

\end{document}